\documentclass[prl,twocolumn,amssymb]{revtex4}
\usepackage{graphicx}

\begin{document}

\title{Correlated vs Ferromagnetic State in Repulsively Interacting Two-Component Fermi Gases }
\author{Hui Zhai}
\affiliation{Institute for Advanced Study, Tsinghua University, Beijing, China, 100084}
\date{\today}
\begin{abstract}
Whether a spin-$1/2$ Fermi gas will become ferromagnetic as the strength of repulsive interaction increases is a long-standing controversial issue. Recently this problem has been studied experimentally by Jo {\it et al}, Science, 325, 1521 (2009) in which the authors claim a ferromagnetic transition is observed. This work is to point out the results of this experiment can not distinguish whether the system is in a ferromagnetic state or in a non-magnetic but strongly short-range correlated state. A conclusive experimental demonstration of ferromagnetism relies on the observation of ferromagnetic domains.
\end{abstract}
\maketitle

Itinerant ferromagnetism is a common phenomenon in nature, but not yet well understood. Rigorous examples of itinerant ferromagnetic ground state have only been obtained for a few specific cases. For instance, Nagaoka shows that for infinite strong repulsive interaction, in a bipartite lattice the ground state is ferromagnetic if one hole is doped into a half-filled system \cite{Nagaoka}. Lieb shows for a half filled bipartite lattice, the ground state of repulsively interacting fermions has non-zero spin if the number of total lattice site of each sub-lattice is not equal \cite{Lieb}. Mielke \cite{Mielke} and Tasaki \cite{Tasaki} propose a class of models whose single particle ground states have degeneracy, and show they become ferromagnetic with repulsive interactions. However, there is no conclusive results for a generic dispersion and filling number.

Stoner considered spin-$1/2$ fermions with short range interactions, spin polarization can lower the interaction energy since two spin align fermions will not interact due to the Pauli exclusion principle, while it costs the kinetic energy. With Hatree-Fock approximation, one can conclude that when $U\mathcal{N}(E_{\text{F}})>1$ there exists a second-order ferromagnetic phase transition \cite{Stoner, Pethick}, where $U$ is the interaction strength and $\mathcal{N}(E_{\text{F}})$ is the density-of-state nearby the Fermi surface. This is known as Stoner criteria. For $s$-wave scattering, this condition corresponds to $k_{\text{F}}a_{\text{s}}>\pi/2$, where $k_{\text{F}}$ is the Fermi momentum, and $a_{\text{s}}$ is the s-wave scattering length. Higher order perturbation of interactions will lower the critical value of $k_{\text{F}}a_{\text{s}}$, and may change the transition to first order \cite{higher}. 

Many authors have proposed to study itinerant ferromagnetism transition using two-component Fermi gases where $a_{\text{s}}$ can be tuned by Feshbach resonance  \cite{Macdonald}. Based on the physical picture above, in a trapped system one should observe non-monotonic dependence of the kinetic energy with the increase of $a_{\text{s}}$, namely, the kinetic energy shall first decrease before ferromagnetic transition due to the expansion of the cloud, and then increase after the transition. The inelastic collision rate shall first increase and then decrease as different components begin to separate spatially \cite{LeBlanc}. Recently, a beautiful experiment by Jo {\it et al} \cite{Ketterle} have observed all these monotonic features, and the agreement between experiment and ferromagnetic theory \cite{LeBlanc,Simons} leads to the claim that this has shown experimentally a ferromagnetic transition in continuum without particular requirement of lattice and band structure \cite{Ketterle}.

However, the itinerant ferromagnetic issue is in fact more complicated than this. The question is, whether spin polarization is the only way to reduce interaction energy. The answer is {\it no}. In the content of Hubbard model, Gutzwiller constructed his famous projected wave function as $\prod_{i}(1-\eta n_{i\uparrow}n_{i\downarrow})|\Psi_0\rangle$, where $|\Psi_0\rangle$ is free fermion Fermi sea, and $i$ is the index of the lattice site. The projection operator $\prod_{i}(1-\eta n_{i\uparrow}n_{i\downarrow})$ ($\eta >0$) suppresses the probability of having two fermions at the same lattice site, and consequently reduces on-site interaction energy \cite{Gutzwiller}. This state is non-magnetic if $|\Psi_0\rangle$ is chosen as non-magnetic state. Hereafter we shall call this state ``correlated state" to distinguish it from ``ferromagnetic state". Nevertheless, we shall note this state is not an exotic state but still a Fermi liquid state, we use the term ``correlated state" in the sense that the projection operator introduces strong short-range correlation into this state. In continuum, a Jastrow factor can play the role of the projection operator. 

In the Hubbard model, using the projected wave function as a variational wave function, Gutzwiller shows that at low-density, the correlated state has lower energy than a ferromagnetic state \cite{Gutzwiller}. An alternative view is that the short-range correlation, which has been ignored in the Hatree-Fock and perturbation treatment, will significantly renormalize down the interaction. Kanamori argued that the up-bound of the effective interaction should correspond to the kinetic cost to put a node in the wave-function where two fermions overlap, which should alway be finite even when bare interaction goes to infinite, and he also argued that the renormalized interaction is not sufficient for ferromagnetic transition at low density \cite{Kanamori}, which is supported by some later calculations \cite{no-ferr}.

In short, the key of the itinerant ferromagnetism problem is whether the system will choose spin polarization or building up short-range correlation to reduce interaction energy as the strength of interaction increases. The advantage of cold atom is to provide an opportunity for a direct quantum simulation of the Stoner model, and hopefully can settle the issue of itinerant ferromagnetism experimentally. So the question comes to whether the experiment of Ref. \cite{Ketterle} has conclusively settled the issue. The answer is {\it no}. The purpose of this Rapid Communication is to point out a non-magnetic ``correlated" state can explain the main observation of Ref. \cite{Ketterle} equally well as a ferromagnetic state, in another word, from the existing experimental results, it is very hard to distinguish whether the system is in a ``correlated" state or in a ferromagnetic state. Further experimental efforts are required to distinguish them.

{\it Equation-of-state for a ``correlated" state.} Let us first consider two-component fermions in free space (without optical lattice and harmonic trap), the Hamiltonian is given by
\begin{equation}
\mathcal{H}=\sum\limits_{{\bf r^{\sigma}_{i}},\sigma=\uparrow,\downarrow}-\frac{\hbar^2\nabla^2_i}{2m}+\sum\limits_{{\bf r^{\uparrow}_i},{\bf r^{\downarrow}_j}} v({\bf r^{\uparrow}_i-r^{\downarrow}_j})
\end{equation}
where $v({\bf r^\uparrow_i-r^{\downarrow}_j})$ is a short-range pairwise interacting potential. For a non-polarized free Fermi sea $|\Psi_0\rangle=\text{Det}(e^{i{\bf k_i} {\bf r^\uparrow_i}})\text{Det}(e^{i {\bf k_j}{\bf r^\downarrow_j}})$, the kinetic energy of each component is given by $E^0_{\text{kin}}=3E_{\text{F}}n/5$, where $E_{\text{F}}=\hbar^2 k^2_{\text{F}}/(2m)$, $n$ is the density of each component, and $n=k^3_{\text{F}}/(6\pi^2)$. For a Fermi sea, the interaction energy is proportional to the  ${\bf k}=0$ Fourier component of $v({\bf r})$ (denoted by $v_0$), i.e. 
\begin{equation}
E^0_{\text{int}}=\langle \Psi_0|\sum_{{\bf r^{\uparrow}_i},{\bf r^{\downarrow}_j}} v({\bf r^{\uparrow}_i-r^{\downarrow}_j})|\Psi_0\rangle=v_{0}n^2.
\end{equation} 
Away from a Feshbach resonance, 
$E^0_{\text{int}}=4\pi\hbar^2 a_{\text{s}}n^2/m$.

Now we consider Gutzwiller's projected wave function in continuum $|\Psi\rangle=\mathcal{P}|\Psi_0\rangle$ as a class of {\it varational} states.
With the projection operator, the probability of having two spin-opposite fermions closely changes from $n^2$ to $(1-g)n^2$, and the interaction energy decreases if $g>0$ and increases if $g<0$, thus the interaction energy shall linearly depend on the ``projection strength" $g$ as
\begin{equation}
E_{\text{int}}=\frac{4\pi\hbar^2 a_{\text{s}}}{m}n^2(1-g)\label{Eint}
\end{equation}
By dimension analysis the kinetic energy shall be of the form
\begin{equation}
E_{\text{kin}}=\langle \Psi|\sum\limits_{{\bf r^\sigma_i},\sigma}-\frac{\hbar^2\nabla^2_i}{2m}|\Psi\rangle=\frac{3}{5}E_{\text{F}}nw(g),
\end{equation}
where $w(g)$ is a dimensionless function of $g$. There are some simple properties of $w(g)$ one can make use of. For $g=0$, there is no projection and the free Fermi sea is the state that minimizes the kinetic energy, thus $w(0)=2$. If $g\neq 0$, both positive and negative $g$ will lead to the increase of the kinetic energy, thus $g=0$ is the minimum of $w(g)$, namely, $\partial w(g)/\partial g|_{g=0}=0$. Hence, up to the second order of $g$, one has the form
\begin{equation}
E_{\text{kin}}=\frac{3}{5}E_{\text{F}}n(2+\alpha g^2)=\frac{4\pi\hbar^2}{m}a_0n^{5/3}(2+\alpha g^2)\label{Ekin}
\end{equation}
where $\alpha>0$, and $a_0=3(6\pi^2)^{2/3}/(40\pi)=0.36$. We shall now stress that the purpose of this work is neither to rigorously derive this equation-of-state and calculate the number of $\alpha$, nor to prove theoretically that this state can energetically do better than a ferromagnetic state. Instead, we shall take Eq. \ref{Eint} and Eq. \ref{Ekin} together as a simple ``{\it phenomenological} " equation-of-state for this class correlated state, and the key of work is to point out the general behavior of this correlated state in trap, which does not depend on the specific value of $\alpha$, and hereafter we shall use $\alpha$ as an unspecified parameter.

\begin{figure}[bp]
\includegraphics[scale=1.0]
{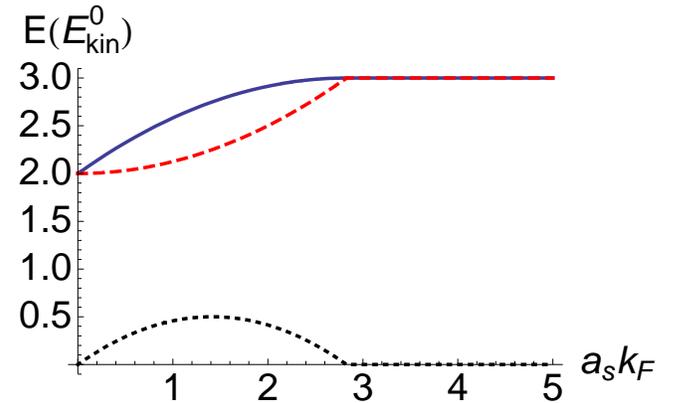}
\caption{(Color online) The total energy (blue solid line), the kinetic energy (red dashed line) and the interaction energy (black dotted line) of the ``correlated state" as a function of $a_{\text{s}}k_{\text{F}}$. For this plot we set $\alpha=1$. $E^0_{\text{kin}}$ is the kinetic energy of a free Fermi sea $\hbar^2 k^2_{\text{F}}/(2m)$. \label{EoS}}
\end{figure}

For a given density $n$ and $a_{\text{s}}$, one shall first minimize the free energy with respect to $g$. For $a_{\text{s}}n^{1/3}\leq 2\alpha a_0$, $g=a_{\text{s}}n^{1/3}/(2\alpha a_0)$. In this regime, 
\begin{eqnarray}
&&E_{\text{kin}}=\frac{4\pi\hbar^2}{m}\left[a_0n^{5/3}\left(2+\frac{a^2_{s}n^{2/3}}{4a^2_0\alpha}\right)\right]\\
&&E_{\text{int}}=\frac{4\pi\hbar^2}{m}\left[a_{\text{s}}n^2\left(1-\frac{a_{s}n^{1/3}}{2a_0\alpha}\right)\right]
\end{eqnarray}
the total energy 
\begin{equation}
E_{\text{tot}}=\frac{4\pi\hbar^2}{m}\left[2 a_0 n^{5/3}+a_{\text{s}}n^2-\frac{a^2_{\text{s}}n^{7/3}}{4a_0\alpha}\right]
\end{equation}
and the chemical potential 
\begin{equation}
\mu=\frac{4\pi\hbar^2}{m}\left[\frac{5a_0 n^{2/3}}{3}+a_{\text{s}}n-\frac{7 a^2_{\text{s}} n^{4/3}}{24 a_0\alpha}\right].
\end{equation}
For $a_{\text{s}}n^{1/3}> 2\alpha a_0$, $g=1$. In this regime,
\begin{equation}
E_{\text{kin}}=\frac{4\pi\hbar^2}{m}\left[a_0 n^{5/3}(2+\alpha)\right],
\end{equation}
and $E_{\text{int}}=0$, the total energy 
\begin{equation}
E_{\text{tot}}=\frac{4\pi\hbar^2}{m}\left[a_0 n^{5/3}(2+\alpha)\right]
\end{equation}
and the chemical potential
\begin{equation}
\mu=\frac{4\pi\hbar^2}{m}\left[\frac{5a_0 n^{2/3}}{6}(2+\alpha)\right].
\end{equation} 
The kinetic, interaction and total energy (in unit of $E^0_{\text{kin}}$) as a function of $a_{\text{s}}k_{\text{F}}$ are illustrated in Fig. \ref{EoS}. When $g=1$ at very large $k_{\text{F}}a_{\text{s}}$, the energy of a correlated state is lower than a fully polarized ferromagnetic state if $\alpha<2^{5/3}-2\sim 1.17$.

\begin{figure}[bp]
\includegraphics[scale=0.48]
{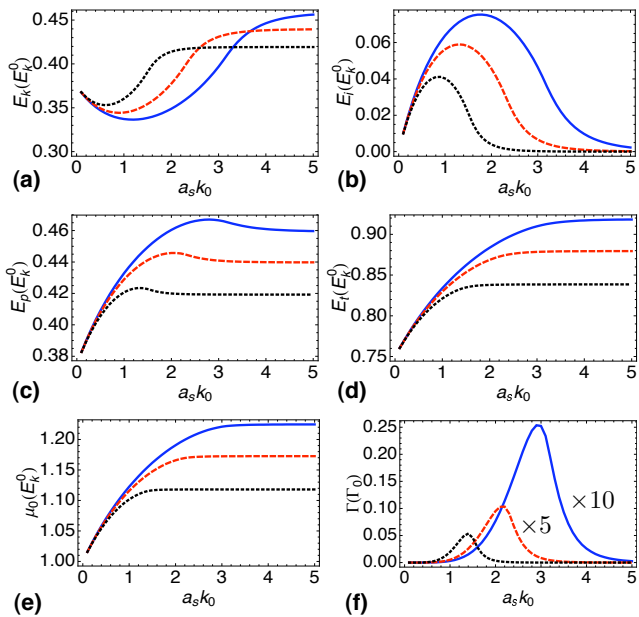}
\caption{(Color online) From (a)-(e) is the kinetic energy per particle $E_{\text{k}}=\mathcal{E}_{\text{kin}}/(2N)$, the interaction energy per particle $E_{\text{i}}=\mathcal{E}_{\text{int}}/(2N)$, the potential energy per particle $E_{\text{p}}=\mathcal{E}_{\text{pot}}/(2N)$, the total energy per particle $E_{\text{t}}=\mathcal{E}_{\text{tot}}/(2N)$ and the chemical potential $\mu_0$ as a function of $a_{\text{s}}k_0$. $k_0$ is the Fermi momentum for free Fermi gas at the center of the trap, and the energy unit is taken as $E^0_{k}=\hbar^2 k^2_0/(2m)$. (f) the three-body loss rate as a function of $a_{\text{s}}k_0$. $\alpha=1$ for blue solid line, $\alpha=0.75$ for red dashed line and $\alpha=0.5$ for black dotted line.  \label{trapped}}
\end{figure}

{\it Trapped System.} From the discussion above, we have obtained the relation $\mu(n,a_{\text{s}})$. For a given $a_{\text{s}}$, one can invert this relation to obtain $n(\mu,a_{\text{s}})$. Considering the harmonic trapping potential $V_{\text{trap}}({\bf r})=(m\omega_{\perp}(x^2+y^2)+m\omega_z z^2)/2$, we shall use local density approximation to replace $\mu$ with $\mu_0-V_{\text{trap}}({\bf r})$ and by solving the total number of particle constraint $\int d^3{\bf r}n(\mu_0-V({\bf r}),a_{\text{s}})=N$, one can obtain $\mu_0(N,a_{\text{s}})$. Then the local fermion density is given by $n({\bf r})=n(\mu_0(N,a_{\text{s}})-V({\bf r}),a_{\text{s}})$. Using the expressions for kinetic and interaction energy density discussed above, one can compute the total kinetic and interaction energy as $\mathcal{E}_{\text{kin/int}}=\int d^3{\bf r}E_{\text{kin/int}}(n({\bf r}), a_{\text{s}})$, and the potential energy is given by $\mathcal{E}_{\text{pot}}=\int d^3{\bf r}V_{\text{trap}}({\bf r})n({\bf r})$, and the total energy is $\mathcal{E}_{\text{tot}}=\mathcal{E}_{\text{kin}}+\mathcal{E}_{\text{int}}+\mathcal{E}_{\text{pot}}$. The loss rate is computed in a very phenomenological way as $\Gamma=2\Gamma_0(a_{\text{s}}k_0)^6\int d^3{\bf r} n^3({\bf r})(1-g(n({\bf r}))$ \cite{loss}.

As shown in Fig. \ref{EoS}, for a uniform system the kinetic energy for a correlated state monotonically increases for any $a_{\text{s}}>0$. To show whether for small $a_{\text{s}}$ the kinetic energy will first decrease with the increase of $a_{\text{s}}$ in a trapped system, we shall note 
\begin{equation}
\frac{\partial\mathcal{E}_{\text{kin}}}{\partial a_{\text{s}}}=\int d^3{\bf r} \left(\frac{\partial E_{\text{kin}}}{\partial n}\frac{\partial n}{\partial \mu}\frac{\partial\mu_0}{\partial a_{\text{s}}}+\frac{\partial E_{\text{kin}}}{\partial a_{\text{s}}}\right).
\end{equation} 
The first term is negative and the second is positive. It is important to note that when $a_{\text{s}}\rightarrow 0$ the first term does not vanish while the second term does, since $\mu_0$ linearly depends on $a_{\text{s}}$ while $E_{\text{kin}}$ quadratically depends on $a_{\text{s}}$, therefore the first term is always dominative in small $a_{\text{s}}$, which gives $\partial \mathcal{E}_{\text{kin}}/\partial a_{\text{s}}<0$, and leads to a non-monotonic behavior of kinetic energy.

We consider the experimental condition as Ref. \cite{Ketterle}, i.e. $N=6.5\times 10^5$ and $\omega_z/\omega_\perp=7/30$. The results are shown in Fig. \ref{trapped}. Comparing them with the prediction of a ferromagnetic state, for instance, Fig 1 and 2 of Ref. \cite{LeBlanc} and Fig. 2 of Ref. \cite{Simons}, they display similar non-monotonic behavior and also qualitatively agree with the observation of Ref. \cite{Ketterle}. This leads to the main point of this work, that is, the non-monotonic behavior observed in Ref. \cite{Ketterle} is not sufficient to distinguish a ferromagnetic state from a non-magnetic correlated state, and thus not conclusive for making the claim of ferromagnetic transition. We emphasize that despite of the similar non-monotonic behavior, there is no phase transition in this scenario. In fact, the suppression of interaction energy and the inelastic collision rate due to correlation is not surprising in strongly interacting systems. Quantum Hall effect and the Tonk gas of one-dimensional bosons are two of the examples. Suppression of the three-body recombination rate has been observed in one-dimensional Bose gas as it approaches the Tonk gas regime \cite{Porto}.

{\it Discussions.} There are a few points we would like to comment on before ending. First, there are some quantitative differences between the results of Fig. \ref{trapped} and that from a ferromagnetic theory (for instance, Fig 1 of Ref. \cite{LeBlanc}). In Fig. \ref{trapped}, the extreme of kinetic energy, potential energy and the loss rate are not very close, while they are very close in the ferromagnetic theory prediction. And there is no maximum in the chemical potential (i.e. cloud size) plot of Fig \ref{trapped}(e). However, both calculation above and the theoretical work of Ref. \cite{Macdonald,LeBlanc,Simons} are not quantitatively correct. The important effect of Feshbach resonance and unitary limit of the repulsive interaction is not taken into account. For instance, the Hatree-Fock energy of a free-Fermi gas is taken as linearly increasing with $a_{\text{s}}$, while the accurate Hatree-Fock energy should be smaller and saturates at large $a_{\text{s}}$. The resonance physics has to be taken into account seriously for making a quantitative comparison between theory and experiments, for instance, the value $k_{\text{F}}a_{\text{s}}$ of kinetic energy turning point, and for constructing a correct microscopic Fermi liquid theory. And for the correlated state, the correction should be treated more seriously rather than the phenomenological way presented above, for instance, by quantum Monte Carlo simulation. It remains to be seen whether these quantitative difference between the prediction of two scenarios can be used to distinguish these two states, when a more careful analysis in the theory is done. We leave this for follow up works.

Secondly, a conclusive experimental evidence of ferromagnetism is the observation of ferromagnetic domains. Ref. \cite{Ketterle} fails to observe the ferromagnetic domains. They attribute this reason to short lifetime that prevents the system to reach equilibrium. However, one should notice that this system is the same as what has been used to study BEC-BCS crossover before. Maybe there is some particular physics reason to believe the relaxation time is particularly long in this case than in the case of BEC-BCS time. If it is the case, the dynamics remains to be explored. 

{\it Acknowledgment:} The author would like to thank Tin-Lun Ho for helpful comments on the manuscript.

\end{document}